\documentclass[apjfonts]{emulateapj}

\slugcomment{Accepted for publication in {\sl The Astrophysical Journal}, August 15, 2005}

\shorttitle{Early-type galaxies in GOODS}
\shortauthors{Ferreras et al.}

\begin{document}

\title{Evolution of field early-type galaxies: The view
       from GOODS/CDFS}

\author{Ignacio Ferreras\altaffilmark{1}, Thorsten Lisker\altaffilmark{2}, 
  C.~Marcella Carollo and Simon~J. Lilly}
\affil{Institut f\"ur Astronomie, ETH Z\"urich, 
  CH-8093 Z\"urich, Switzerland}
\and
\author{Bahram Mobasher}
\affil{Space Telescope Science Institute, 3700 San Martin Drive, 
  Baltimore, MD 21218, USA}
\altaffiltext{1}{Current address: Department of Physics and Astronomy,
University College London, Gower Street, London WC1E 6BT, United Kingdom}
\altaffiltext{2}{Current address: Astronomical Institute, University of Basel, 
Venusstrasse 7, CH-4102 Binningen, Switzerland}

\begin{abstract}
We explore the evolution of field early-type galaxies on a sample
extracted from the ACS images of the Southern GOODS field.  The
galaxies are selected by means of a non-parametric analysis followed
by visual inspection of the candidates with a concentrated surface
brightness distribution.  We furthermore exclude from the final sample
those galaxies which are not consistent with an evolution into the
Kormendy relation between surface brightness and size that is observed
for $z=0$ ellipticals.  The final set -- which comprises $249$
galaxies with a median redshift $z_m=0.71$ -- represents a sample of
early-type systems {\em not} selected with respect to color, with
similar scaling relations as those of bona fide elliptical galaxies.
The distribution of number counts versus apparent magnitude rejects a
constant number density with cosmic time and suggests a substantial
decrease with redshift: $n\propto (1+z)^{-2.5}$. The majority of the
galaxies (78\%) feature passively evolving old stellar populations.
One third of those in the upper half of the redshift distribution have
blue colors, in contrast to only 10\% in the lower redshift
subsample. An adaptive binning of the color maps using an optimal
Voronoi tessellation is performed to explore the internal color
distribution. We find that the red and blue early-type galaxies in our
sample have distinct behavior with respect to the \emph{color
gradients}, so that most blue galaxies feature blue cores whereas most
of the red early-types are passively evolving stellar populations with
red cores, i.e. similar systems to local early-type
galaxies. Furthermore, the color gradients and scatter do not evolve
with redshift and are compatible with the observations at $z=0$
assuming a radial dependence of the metallicity within each
galaxy. Significant gradients in the stellar age are readily ruled
out. This work emphasizes the need for a careful sample selection, as
we found that most of those galaxies which were visually classified as
candidate early types -- but then rejected based on the Kormendy
relation -- feature blue colors characteristic of recent star
formation.
\end{abstract}

\keywords{galaxies: elliptical and lenticular, cD - galaxies: evolution - 
galaxies: stellar content}

\section{Introduction}

The formation of early-type galaxies poses one of the most intriguing 
riddles of astrophysics. Their stellar populations reveal an early formation
process, with no major star formation at redshifts $z\lesssim 2$
(e.g. Bower, Lucey \& Ellis 1992; Stanford, Eisenhardt \& Dickinson 1998). 
Recent findings of spheroidal galaxies at $z\gtrsim 1$ with a massive 
stellar component ($\gtrsim 10^{11}$~M$_\odot$) already in place 
confirms this early buildup. Furthermore, these red and old galaxies make
a 30\% contribution to the total stellar mass content in the Universe at $z\sim 1-2$
 (Glazebrook et al. 2004; Cimatti et al. 2004).
The tight correlation between color and luminosity and its evolution 
with redshift implies a mass-metallicity relation 
(Kodama \& Arimoto 1997; Kuntschner 2000). 
Furthermore, the observed enhancement of [Mg/Fe] in giant
ellipticals with respect to solar abundance ratios 
(e.g. Carollo, Buson \& Danziger 1993; Carollo \& Danziger 1994a,1994b; 
Trager et al. 2000; Eisenstein et al. 2003) 
implies a very short period of formation which should not be any 
longer than 1~Gyr (e.g. Thomas, Greggio \& Bender 1999).
Hence, the photo-spectroscopic properties of giant early-type 
galaxies consistently reveals an early, intense and brief period 
of star formation inside a deep potential well.

On the other hand, our current knowledge of structure formation 
suggests a hierarchical buildup, so that small galaxies form
first and massive galaxies and clusters of galaxies assemble 
later (e.g. White \& Rees 1978).
Semi-analytical models have combined simulations of structure formation to
simple prescriptions of star formation (e.g. Baugh et al. 1998; 
Kaufmann \& Charlot 1998; Somerville \& Primack  1999) and have come up
with significant achievements in our understanding of many properties
of galaxy formation. However, these models have also been
criticized as being too contrived as many parameters are used and
adjusted in order to fit the observations {\sl a posteriori}.
The current problem of [Mg/Fe] enhancement
in massive ellipticals has posed so far one of the most important
challenges to these models (Thomas 1999). 

Quite often in the literature an 
alternative scenario is invoked, namely the monolithic collapse model, based
on the early attempts of Eggen, Lynden-Bell \& Sandage (1962) at explaining
the formation of our Milky Way. According to this model the Galaxy
formed very early and in a rapid collapse. The extension 
to early-type galaxies assumes a highly efficient formation process so that
most of the stellar populations are formed early, followed by
pure passive evolution. Such a model
is compatible with the stellar populations of early-type galaxies as 
discussed above. However, monolithic collapse does not include any information 
on the constraints that the cosmology imposes on structure formation.

The main drawback in the study of early-type galaxies at
moderate and high redshift lies in the sample selection. The 
outstanding spatial resolution of the Hubble Space Telescope
has enabled these searches in the past few years. The Wide Field
and Planetary Camera~2 (WFPC2) and the more recent Advanced Camera for
Surveys (ACS) allow a resolved analysis of the projected surface
brightness profiles for galaxies with characteristic sizes down 
to $\sim 0.1$~arcsec, which maps into a physical projected size of 
$\sim 1$~kpc at $z=1$ for our chosen cosmology (see below). Deep surveys
such as the Hubble Deep Fields probe deep enough to 
allow a resolved color map of early-type galaxies at moderate and
high redshift (e.g. Tamura \& Ohta 2000; Menanteau, Abraham \& Ellis 2001).

Recent surveys targeting early-type galaxies at moderate and
high redshift (e.g. Franceschini et al. 1998; Schade et al. 1999; 
Stanford et al. 2004) have come up with a common picture, namely that 
these galaxies have most of their stellar populations formed before
$z\gtrsim 2$ in agreement with previous studies.  
No significant differences are found between cluster and field
early-type galaxies (van~Dokkum et al. 2001). However, there is evidence
for a redshift evolution of the size and stellar mass content 
(Fasano et al. 1998; Trujillo \& Aguerri 2004; Bell et al. 2004; 
Fontana et al. 2004).
Furthermore, the resolved color analysis of samples of early-type
galaxies in the HDF (Menanteau et al. 2001) and in the field of the
Tadpole galaxy (Menanteau et al 2004) have revealed that $\sim 20$\% of
the sample of galaxies classified as early-types present blue cores
characteristic of recent or ongoing star formation. Moreover, 
spectroscopy has also revealed [\ion{O}{2}] emission lines in 
a similar fraction of field early-type galaxies (Schade et al. 1999; 
Treu et al. 2002). 

However, all these analyses critically hinge on the criterion used
to determine whether a galaxy belongs to an early-type morphology or not.
There are two main approaches to define such a type: either through its
morphology or its stellar populations. The latter is a fairly straightforward
method to use and indeed has been the main selection criterion of many
surveys (e.g. Bell et al. 2004). However, this method only targets
a predefined set of stellar populations. Ellipticals with a significant
amount of star formation would be misclassified in this way.

A morphological analysis allows us to
compare the same type of objects locally and at high redshift.
In this paper, we start with a non-parametric approach for the classification
(e.g. Abraham et al. 1994). We believe this method is very robust for the faint and small 
galaxies found in high redshift surveys, especially when complemented 
by subsequent visual inspection.
We pay special attention to the sample selection
and throughout this project we question whether the selected objects
are indeed early-type galaxies. The sample is also split with respect to
the observed optical and
NIR colors as classified by Mobasher et al. (2004). Different tests
regarding the color distribution, surface brightness and size will
lead us to reject most of the bluer systems. This is a result with 
important consequences in the interpretation of present and upcoming 
surveys of early-type galaxies.

Throughout this paper we use a concordance $\Lambda CDM$ cosmology with
$\Omega_m=0.3$ and $H_0=70$~km~s$^{-1}$~Mpc$^{-1}$.
For this cosmology, the age of the Universe is $13.5$~Gyr and the look back
times at $z=0.5$ and $z=1$ are $5.0$ and $7.7$~Gyr, respectively. All magnitudes
are given in the AB system.

\section{The data}

Our sample is selected from the v1.0 release of the HST/ACS images of the
GOODS-South field (Giavalisco et al. 2004). These images cover 
a $160$~arcmin$^2$ area towards the Chandra Deep Field South, and 
consist of four deep images through the F435W($B$), F606W($V$), 
F775W($i$) and F850LP($z$) passbands of ACS. The field is divided into 18
sections and the images are drizzled with a pixel size of 30~mas --
chosen to optimize the sampling of the PSF. Each image comes with 
a weight map given as the expected inverse variance per pixel. The
weight maps are used both to assess uncertainties in the direct 
measurements as well as to generate mock images of early-type 
galaxies in order to perform a Monte Carlo analysis of the accuracy 
of the color gradients and scatter (see appendix). 

Central to this project is a redshift estimate and a photometric 
classification. We use the photometric redshift catalog
of Mobasher et al. (2004) as well as the spectroscopic follow-up from 
the VLT/FORS2 (Vanzella et al. 2005) and VLT/VIMOS (Le~F\`evre et al. 2004) 
redshift surveys. The photometric redshifts of
Mobasher et al. (2004) have an RMS scatter with respect to the 
spectroscopic measurements of $\sigma(\Delta z/1+z)\leq 0.1$.
The ``photometric'' type is a byproduct of the classification and corresponds
to the template that gives the best fit to the photometric data. This
type ranges from 1 (early-type galaxy) to 6 (starburst) and is interpolated
in steps of $1/3$. Our classification
does \emph{not} reject candidates based on this photometric type, i.e.
our sample is not biassed in favor of red galaxies. However,
in order to explore the connection between the morphological and photometric
classification, we decided to separate the final sample (see \S3 and \S4) 
with respect to photometric types, so that galaxies with a type $t\leq 1\frac{1}{3}$
are termed red galaxies and $t>1\frac{1}{3}$ are blue galaxies.

\begin{figure}
\epsscale{.80}
\plotone{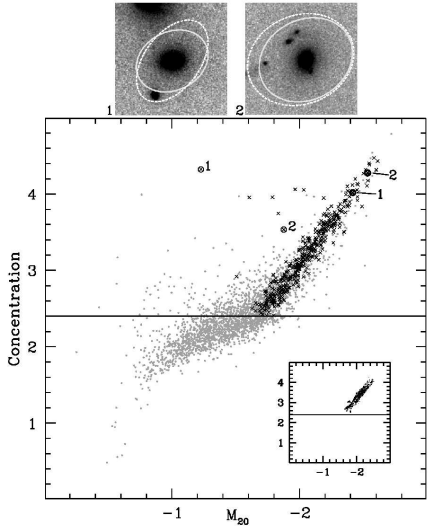}
\caption{The $C$-$M_{20}$-distribution of all galaxies with
    $i_{\rm AB}\le 24$: visually selected early-types (black crosses)
    and the rest (grey dots). The F775W images of two galaxies are
    shown above the diagram, and their positions before and after the
    manual replacement of neighboring sources are labeled. Both images show
    the Petrosian aperture before (dashed line) and after the
    replacement. (Note that this replacement is not applied to the images
    shown here, i.e.~all sources are still there.) The inset shows the
    $C$-$M_{20}$-distribution of simulated galaxies (see text).
\label{fig:morpho}}
\end{figure}

\section{Morphological classification}

In order to create a first object catalog, source detection was
performed in the F775W band using the Source Extractor Software
(hereafter SExtractor, Bertin \& Arnouts 1996). 
Even though the reddest band would be the obvious choice
when exploring a large range of redshifts, we discarded the F850LP
bandpass as the images through this filter are not as deep as in the
F775W passband. From SExtractor, we obtained segmentation images --
which define the pixels associated to each object -- for a signal
threshold of $1\,\sigma$. We used these as input for the computation of
an elliptical aperture for each source, with the ellipticity and
position angle derived from the image moments (e.g.~Abraham et
al.~1994), and the semimajor axis (SMA) chosen to be 1.0
'Petrosian SMA'. The latter means that we applied the concept of the
Petrosian radius (Petrosian 1976). It is chosen so that the ratio 
between the surface brightness at that radius ($r_p$) and 
the average one inside $r_p$ is fixed at a constant value:
$\eta\equiv\Sigma (r_p)/\langle \Sigma (\leq r_p)\rangle = 0.2$ 
(see e.g. Lotz et al.~2004).
We used elliptical apertures defined by the 
image moments of the initial segmentation image. Choosing elliptical
instead of circular apertures guarantees that we do not unnecessarily
include a large number of noise-dominated pixels, which is important for
obtaining reliable morphological parameters.

Although previous studies preferentially used apertures
of 1.5 Petrosian SMA (e.g.~Lotz et al.~2004), it seemed more useful for the
identification of concentrated objects to choose 1.0 Petrosian SMA: first, we
avoid for all types of galaxies the inclusion of noise-dominated pixels into the
aperture, which overall allows a more robust classification, and second, we
lower the amount of contamination by nearby sources. Due to this
approach, we miss a certain fraction of the galaxy light (see e.g.
Blanton et al.~2001), but since this is consistently applied
throughout the paper, it does not prevent us from benefiting from the
advantages named above.

Still, in order to obtain
a useful Petrosian SMA, it was inevitable to replace all sources -- as defined
by the segmentation image -- in the vicinity of the current object by random
gaussian noise matching the background properties, following the approach of
Conselice et al.~(2005). These neighbors would confuse not only the
Petrosian SMA, but also the morphological parameters derived
subsequently. Note that the fraction of objects whose aperture
would be significantly affected by one or more nearby sources is 13\%.
In addition, we performed a proper background subtraction as described 
in \S\S6.2.

Magnitudes were derived using the final aperture, and all objects 
with $i_{\rm AB}>24.0$ were excluded, with $2,837$ galaxies remaining.
We then computed model-independent morphological parameters on that
aperture, namely central concentration (C, Bershady et al.~2000), 
asymmetry (A, Conselice et al.~2000), the Gini coefficient (G, Lotz et
al.~2004), and M$_{20}$ (Lotz et al.~2004). Our goal was to find an
appropriate selection criterion using these parameters,
which on the one hand should guarantee inclusion of all (potential)
spheroids, but on the other hand should exclude the maximum number of
non-spheroids.

\begin{figure}
\plotone{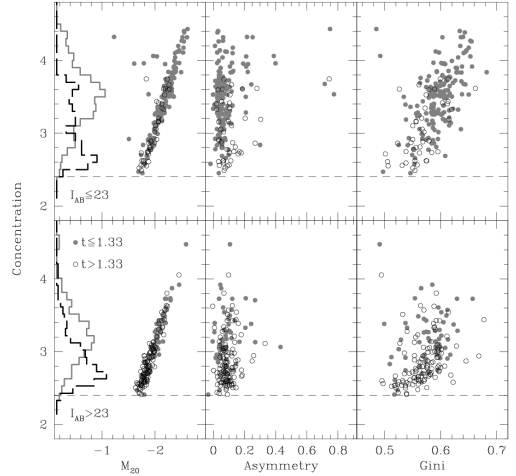}
\caption{The morphological estimates of concentration, asymmetry, 
$M_{20}$, and the Gini coefficient are shown for our final sample, 
split into bright ({\sl top}) 
and faint ({\sl bottom}) galaxies, divided roughly at the value of 
the median ($i_{\rm AB}=22.9$). Throughout the paper we divide the sample
into red  (filled circles; $t\leq 1\frac{1}{3}$)
and blue galaxies (hollow circles; $t>1\frac{1}{3}$). The dashed line 
signals the main selection criterion, namely $C>2.40$. The histograms
of the red and blue subsamples are shown in the left
panels.
\label{fig:cas}}
\end{figure}

We show in Figure \ref{fig:morpho} that basically all galaxies follow a
linear relation in the C-M$_{20}$-plane, which changes its slope at
about C=2.5, M$_{20}=-1.7$. From the definition of the two parameters,
we deduce that this marks the transition between bulge-dominated and
disk-dominated objects: M$_{20}$ traces the spatial distribution of the
brightest 20\% of all object pixels, \emph{no matter where} they are
located, while C is sensitive to the \emph{central} brightest 20\%. It
follows that, at the point when the disk -- or more general: the outer
part of a galaxy -- starts contributing to the \emph{total} 20\%, the
M$_{20}$ increases fast, represented by the lower part of the
sequence. We therefore decided to exclude all galaxies with a
concentration below $C=2.4$, since they lie
below the transition point, allowing for an uncertainty of 0.1. This
leaves us with a subsample of $1,540$ candidate early type galaxies, or
54\% of the full sample.

Four of us (I.F., T.L., C.M.C., S.J.L.) subsequently
performed independent visual classification for the
whole candidate sample, inspecting all four GOODS/ACS bands. 
A galaxy was included in the sample if three people classified it as E/S0.
We selected $372$ galaxies from the list of $1,540$ candidates. Furthermore,
we revisited the sub-sample of $274$ galaxies which were classified by only 
1 or 2 of us as early types. In this second pass, a galaxy was included in the
final list if 3 of us independently decided favorably. 
This increased the list to an
additional eight members, for a total of $380$ early-type galaxies, or
25\% of the candidates (see \S6 for a discussion of this small
fraction). This sample 
was cross-correlated with the photometric redshift catalog of Mobasher 
et al.~(2004) and the GOODS ACS catalog v1.1 and constitutes our
visually classified sample, shown as black crosses in Figure \ref{fig:morpho}.

Nine of these objects do not fall on the linear relation, but instead show a
higher $M_{20}$ value. Two of these are contaminated by
bright artifacts that frequently occur on the edges of the GOODS ACS v1.0
images, whereas the other seven have one or more very close or even
connected neighbors. These contaminants had
not been deblended in the SExtractor segmentation images. However, by
manually replacing the contaminants with noise where they are not
too close to the main object, we show in Figure \ref{fig:morpho} that 
the parameters of all our objects fall onto the observed
$C-M_{20}$-relation: the objects labeled 1 and 2 first occupy the
left position marked with the corresponding number, but after the
manual replacement, they switch to the marked positions on the right.

When we neglect objects that do not follow the $C-M_{20}$ relation, we can
apply cuts around our sample in the multidimensional parameter space, in order
to see whether a more efficient early-type candidate selection would be
possible. By choosing $C\geq 2.40$, $M_{20}\leq -1.60$, $G\geq 0.48$, $A\leq
0.35$, we end up with $1,099$ candidate early types, thus increasing the fraction of
visually classified early types to 34\%. 
The visually classified sample is shown in figure~\ref{fig:cas} as a function
of the morphological parameters explored in this paper. Red and blue galaxies are
coded as grey and hollow circles, respectively (see below). The
panels on the left also show the histogram of the concentrations for 
red and blue galaxies as a solid and dashed line, respectively. 

Several authors have advertised the use of model-independent morphological 
parameters for galaxy classification (e.g. Lotz et al. 2004, Conselice 2003). 
Therefore, it might appear surprising that our morphological pre-selection 
was not able to exclude more than 46\% of the initial sample, and that only 
one quarter of the so selected objects passed the subsequent visual 
classification. There is, however, a major difference in the application of 
those parameters between our study and the ones named above: while the latter 
aimed at a \emph{statistical} segregation of galaxy types, it has been our 
goal to obtain a \emph{complete} sample of early type galaxies. If we instead 
wanted to present only a rough dividing line of early and late type objects, 
a possible choice would be $C>3$, since this includes 75\% of our final sample 
while excluding 90\% of other galaxies. However, given the range of simulated 
early-type objects shown in Fig. 1 as well as our successful selection of a 
complete sample (see \S4), it is clear that the goal of completeness 
can only be achieved through a selection process like the one presented in 
this paper.

\begin{figure}
\plotone{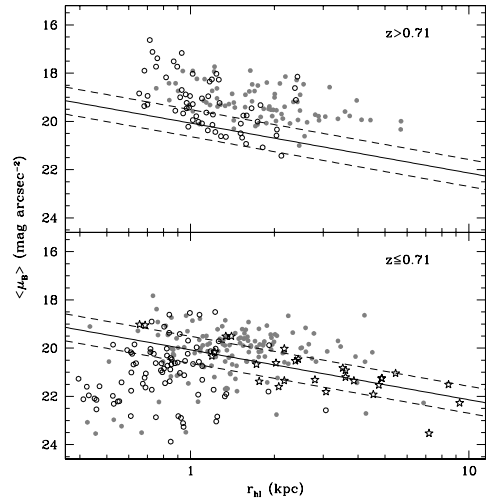}
\caption{Rest frame $B$-band Kormendy relation. Our sample is subdivided
into red (grey dots; $\leq 1\frac{1}{3}$) and blue galaxies (hollow circles). 
The K correction is performed using the spectral templates used
in the determination of the photometric types. The K correction is 
chosen as the smallest one between 
the observed $V$ band or $I$ band and the rest frame $B$ band. 
The stars give the Coma cluster 
relation from J\o rgensen, Franx \& Kj\ae rgaard (1995), whose best 
linear fit is given by the solid line and dashed lines (biweight scatter). 
\label{fig:kor}}
\end{figure}

\section{Photometric classification}

Our preliminary visual classification selects objects with a
concentrated surface brightness profile. Bright nuclei or 
knots of star formation -- when viewed at lower signal-to-noise 
ratios and with apparent sizes close to the resolution limit 
of the ACS images -- can be classified as early-type galaxies.
We decided to impose a further constraint on our sample, based
on the observed scaling relations of early-type galaxies.
The way we choose to ascertain the connection between our sample and
local elliptical galaxies is through the Kormendy relation (Kormendy 1977).
This is one of the projections of the Fundamental Plane
and correlates the size and the rest frame average surface brightness 
inside the effective radius. The Kormendy relation is a
powerful and ``cheap'' technique since it only involves photometry,
and it is found to hold at $z>0$ (La~Barbera et al. 2003).
Figure~\ref{fig:kor} shows the Kormendy relation for our sample, separated
into a low- ({\sl bottom}) and high-redshift subsample ({\sl top}).
We show the surface
brightness in the rest frame $B$ band, which requires a $K$ correction
from a nearby observed filter.
We have applied the photometric type in order to choose the template SED
needed to estimate the $K$ correction. The grey and hollow 
circles correspond to red and blue galaxies, respectively.
The stars are Coma cluster galaxies (J\o rgensen, Franx \& Kj\ae rgaard 
1995) and the lines represent the best fit and scatter. 

\begin{figure}
\plotone{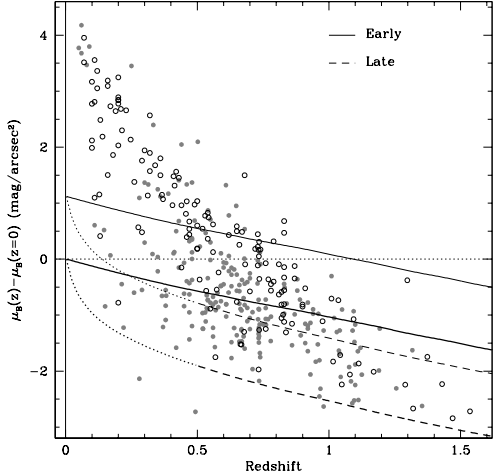}
\caption{Difference between the average surface brightness (at the
observed redshift of each galaxy in the rest frame $B$ band) and the
local measurement expected from the Kormendy relation, as a function
of redshift.  The evolution of two star formation histories with a
formation redshift $z_F=5$ is shown. The thick solid line corresponds
to an exponentially decaying star formation rate with timescale
$\tau_{SF} =0.5$~Gyr (labelled Early).  The thick dashed line assumes
a constant star formation rate from $z_F=5$ to the observed redshift
followed by a truncation of the SFR (labelled Late). The thin lines
are equivalent to the previous star formation histories with an offset
of $2\times$ scatter found in local galaxies. Notice the dashed
line becomes dotted at $z<0.5$. At these redshifts, such models would
be in conflict with the photometry of local early-type
galaxies.\label{fig:korz}}
\end{figure}

The figure shows a remarkable segregation with respect to color.
Notice the large number of blue galaxies in the bottom panel which
fall below the dashed lines that mark the local relation.
Passive fading of the stellar populations implies that 
these galaxies will evolve with redshift towards fainter surface
brightness, drifting away from the relation found
in present day early-type galaxies.
This suggests that most of the blue
galaxies at $z\leq 0.71$ are not early-type galaxies,
whereas most of the red galaxies plus some of the 
high-redshift blue galaxies at $z>0.71$ would qualify as bona 
fide early-type systems. 

We can thereby select only those galaxies which will evolve into the 
observed local Kormendy relation. Figure~\ref{fig:korz} shows 
the change between
the observed surface brightness and the one corresponding to the Kormendy
relation for a given $r_{hl}$ as a function of redshift. 
The sample is divided into red (grey dots) 
and blue galaxies (hollow dots) as usual.
The thick solid line gives the evolution of an exponentially decaying
star formation rate which started at $z_F=5$ with a timescale 
$\tau_{SF}=0.5$~Gyr, and the dashed line corresponds to a constant SFR
truncated at the observed redshift, both with solar metallicity
(population synthesis models from Bruzual \& Charlot 2003). 
Both formation histories are required to have $\Delta\mu_B=0$
at $z=0$. Notice the dashed line becomes dotted at $z<0.5$ to
represent the fact that an extended star formation history at those
redshifts would be in conflict with the photometry of early-type galaxies.
The thin lines are offsets of the
previous formation histories by $2\times$ the observed scatter
found in the local sample (J\o rgensen et al. 1995). 
The threshold imposed depends on the photometric type, as defined
in Mobasher et al. (2004). We interpolate
between these two extreme SFHs, so that the thin solid line is the
limit for a type=1 (red) galaxy and the thin dashed limit
corresponds to a starburst (type=6). All galaxies which fall above
their corresponding threshold will evolve with age into a region 
outside of the Kormendy relation.
The Kormendy-consistent sample comprises $249$ galaxies, 
i.e. about 70\% of the visually selected sample, with a 
significant rejection of  blue galaxies.
In fact, 90\% of the blue $z\leq 0.71$ subsample 
are rejected, whereas only 34\% are rejected in the higher redshift 
case. In contrast, most of the red systems are included in
the final list: only 12\% of these galaxies are rejected.

\begin{figure}
\plotone{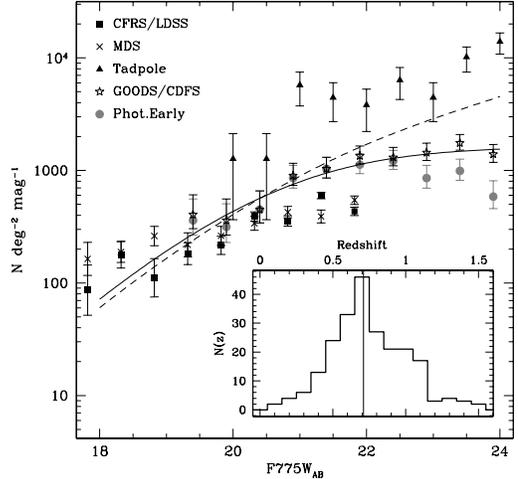}
\caption{Number counts of early-type galaxies in the GOODS/CDFS field
(stars) compared to the CFRS/LDSS redshift surveys (Brinchmann et al. 1998), 
the Medium Deep Survey (MDS; Abraham et al. 1996) and the high-redshift 
field sample 
towards the Tadpole field (Menanteau et al. 2004). 
Our sample is also shown for the subsample of red
early-types ($t\leq 1\frac{1}{3}$; filled circles).
Poisson error bars are shown. The lines give two number 
density predictions using the local luminosity function of E/S0s 
of Marzke et al. (1998). The dashed line assumes a constant comoving
density and includes the photometric evolution of a typical early-type
galaxy (see text for details). The solid line adds to the color evolution 
a change in the comoving number density $n\propto (1+z)^{-2.5}$.
The inset gives the redshift distribution
of our sample, with the median at $z_m=0.71$.
\label{fig:Nm}}
\end{figure}

The rejection process eliminates
50\% of faint galaxies ($i_{\rm AB}>23$) and only 18\% of the
brighter subsample ($i_{\rm AB}\leq 23$). Even though the $i$-band
ACS images of the GOODS field reach a $10\sigma$ limiting magnitude
of $i_{\rm AB}=27.1$ with a $30$~mas sampling and $\sim 100$~mas 
resolution, about one third of the visually classified objects 
must be rejected. A $i_{\rm AB}=24$ limit can be taken as a 
reasonable magnitude limit for the visual classification in these images.
Within that limit it is unlikely that we have missed any early-type galaxy.
First, simulated galaxy images with a S\'ersic surface
brightness profile with indices between $3-4$ and similar magnitudes and sizes
of those in the real sample cover the range $2.5 < C < 4.3$, and thus lie
clearly above our cut of $C=2.4$ (Figure \ref{fig:morpho}). Second, the
majority of galaxies that are 
close to the lower concentration end are faint ($i_{\rm AB}\ge 23$),
small (r$_{\rm hl}\le 0.2$ arcsec), blue objects, suggesting
regions of star formation instead of {\em bona fide} early-type galaxies.
Third, we went back to the photometric redshift catalog and searched for
red galaxies below the concentration threshold imposed
-- since they constitute the majority of our final sample.
We found a total of 30 objects (with $C<2.4$) whose images
clearly do not correspond to an early type morphology.

\begin{figure}
\plotone{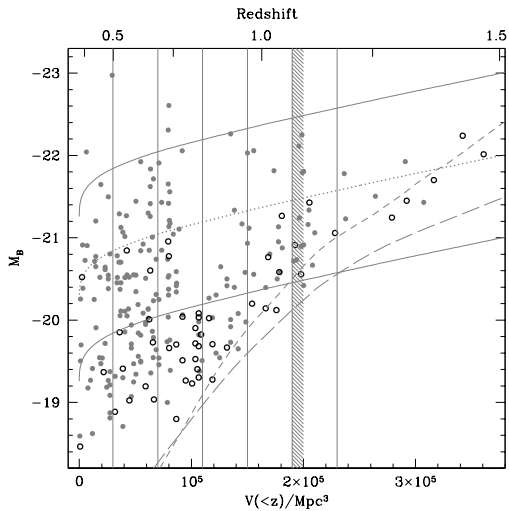}
\caption{Distribution in absolute luminosity as a function of enclosed
  comoving volume. This figure allows us to determine the luminosity
  function of the sample since the number of galaxies inside equal
  segments along the horizontal axis (as shown by the vertical lines)
  corresponds to the number of galaxies inside the same comoving
  volume. The thick dotted line tracks the evolution of a $M_\star$
  galaxy using a star formation history with solar metallicity,
  formation redshift $z_F=5$ and star formation timescale
  $\tau_{SF}=0.5$~Gyr. The thick solid lines represent $M_\star\pm
  1$~mag used to determine a simple luminosity function as shown in
  the next figure.  The thin lines mark the $i_{\rm AB}\leq 24$
  limiting magnitude for a red (short dashed) and blue galaxy (long
  dashed).  The shaded region indicates that the last redshift bin
  starts showing a significant incompleteness for red
  spheroids.\label{fig:PLF}}
\end{figure}

\section {Number counts}
Figure~\ref{fig:Nm} shows the number counts as a function of
apparent magnitude ($i_{\rm AB}$) from our sample (stars), compared to 
the number counts from the CFRS/LDSS survey (squares; 
Brinchmann et al. 1998) and 
the Medium Deep Survey (crosses; Abraham et al. 1996). The sample of
field early-type galaxies in the HST/ACS image of the Tadpole galaxy
is also shown (triangles; Menanteau et al. 2004). Our subsample of
red galaxies is shown as filled dots. The Tadpole sample yields 
a remarkably large number density of objects, with a large
fraction of blue cores\footnote{A transformation had to be
applied to translate their F814W(Vega) into our F775W(AB) photometry. 
It amounts to $\sim 0.7\pm 0.15$~mag, which is still clearly insufficent
to reconcile the number counts}.
We believe this large density is caused by their broader
visual definition of an early-type galaxy. Indeed, a significant fraction
of their candidates  would have been selected by us as late-type systems. 
The dashed line represents a pure luminosity evolution model 
(i.e. fixed comoving density) and uses the $K$ correction corresponding to an 
exponentially decaying star formation rate starting at $z_F=5$ 
with $\tau_{SF} =0.5$~Gyr and solar metallicity. 
In order to determine the number counts we use the local luminosity function 
of early-type galaxies from Marzke et al. (1998) adapted to our cosmology. 
The solid line gives the number counts assuming a
decrease of the comoving number density of galaxies as 
$n\propto (1+z)^{-2.5}$. Notice the drop of the subsample
of red galaxies at $i_{\rm AB}\geq 23$. 
Given that the ACS images have a limiting magnitude that goes $4$
magnitudes fainter than this level, we do not anticipate incompleteness
regarding the apparent magnitudes. However, one could expect incompleteness
because of spatial resolution. Our selection criterion
is robust for galaxies with a $r_{hl}\geq 2$~pixels, which translates
into a projected distance of 0.4~kpc at $z=0.5$ or $0.5$~kpc at $z=1$.
Therefore, we should not expect to miss early-type galaxies because of
the spatial resolution of the ACS images.

The inset in figure~\ref{fig:Nm} gives the redshift distribution of 
our final sample (the vertical line gives the median, $z_m=0.71$),
and features a peak in the number of galaxies around 
$z\sim 0.7$, which is partly caused by the effect of cosmic variance 
given the field of view ($160$~arcmin$^2$; Somerville et al 2004). 
This is a well-known spike (Gilli et al. 2003) 
also seen in K20, a K-selected survey of galaxies 
with a pointing towards the same field (Cimatti et al. 2002).  
The spatial distribution of the galaxies at $z\sim 0.7$
implies a loose structure rather than a cluster.
Despite cosmic variance, most studies, including our own, 
provide consistent number counts. We therefore use our data to estimate
the evolution of the number density of early-type galaxies up to 
$z\sim 1.2$.

Figure~\ref{fig:PLF} shows the $B$-band absolute luminosity with respect 
to the comoving volume out to the observed redshift of each galaxy. 
It  can be interpreted as a luminosity function. The number of
galaxies inside equal segments along the horizontal axis 
gives a comoving number density. The vertical lines show an example of 
redshift ranges delimiting shells with constant comoving volume 
($\Delta V=4\times 10^4$~Mpc$^3$), as used in the next figure. 
The thick dotted line corresponds to a typical $M_\star$ galaxy
and the thick solid lines are offsets $\pm 1$~mag with respect to this value.
The limiting magnitude -- which depends on the choice of star formation history --
is shown as thin short and long dashed lines for a short ($\tau_{SF}=0.5$~Gyr) 
or more extended (8~Gyr) SFH, respectively. Counting the galaxies
within the cells determined by the vertical lines and the
thick lines which track $M_\star$ and $M_\star\pm 1^m$ we can assess
the evolution of the number density of galaxies. 
Notice that the last redshift bin presents a significant imnompleteness
in the detection of red spheroids. We have included a shaded region to
illustrate this point.

\begin{figure}
\plotone{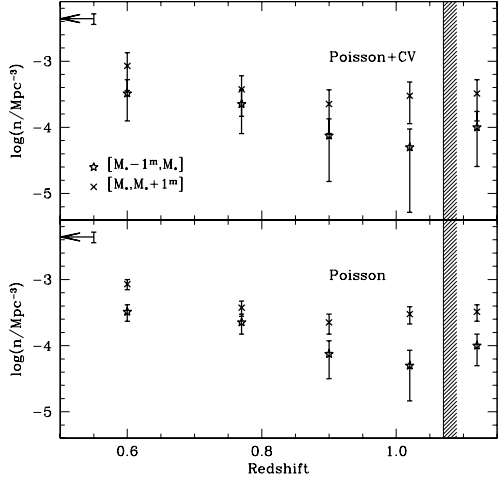}
\caption{The evolution of the luminosity function is shown, according
to the data presented in the previous figure. For each redshift bin we
give the comoving number density for two luminosity bins around
$M_\star$. The bottom panel shows the data along with Poisson error
bars. The top panel includes in quadrature the effect of cosmic
variance according to the models of Somerville et al. (2004). The
shaded area shows the redshift at which the sample is not
complete. The arrow and error bar gives the local number density of
E/S0 from Marzke et al. (1998).\label{fig:CV}}
\end{figure}

Figure~\ref{fig:CV}
shows these counts including simple Poisson error bars (bottom panel)
or a more realistic estimate of the uncertainty by adding in quadrature
the cosmic variance (CV) from the models of Somerville et al. (2004) for a 
correlation function $\xi(r)=(r_0/r)^\gamma$, with $r_0=5h^{-1}$~Mpc 
and $\gamma =1.8$. The stars and crosses represent the bright and faint
luminosity bins, respectively. The shaded region shows the limit at
which we have completeness. The arrow and error bar gives the local
comoving number density of early-type galaxies from Marzke et al. (1998).
Our data show a decrease of a factor $\sim 3$ in the comoving number density
of galaxies {\it with an early-type morphology} in the redshift interval
$z\sim 0.6-1.2$. A survey such as 
COSMOS\footnote{See {\tt http://www.astro.caltech.edu/$\sim$cosmos}}, 
with its $2$~deg$^2$ field of view will be able to assess the scatter 
around this factor due to cosmic variance and large scale structure 
environment.

\section{Resolved color analysis}

The high resolution, good sampling and depth of the HST/ACS GOODS
images enable us to perform a resolved analysis of the color
distribution in our sample of early-types. The 30~mas drizzled pixel
translates into a projected distance of 180~pc at $z=0.5$ and 240~pc
at $z=1.0$.

A good alignment of the two images is crucial, as offsets down to a
fraction of a pixel can generate a fake color gradient -- see Zheng et
al. (2004) for a comprehensive analysis of this issue.  Our images
come from the release v1.0 of the GOODS/ACS images, which were
specifically corrected for geometric distortion. The alignment of the
images have a clipped RMS residual between 1/4 and 1/3 of a
pixel. Nevertheless, we decided to check the alignment by comparing
the centroids of the F606W and F775W images from our detections. Out
of the list of $380$ galaxies from the visually classified sample, we
found a mean offset below $0.02$ pixels, with a RMS below
$0.15$~pixel.  If we use a S\'ersic profile and take the worst case
scenario, i.e. a $0.15$~pixel offset with $r_{\rm hl}=0.2$~arcsec, we
get a maximum color offset below $0.05$~mag. The mean gives offsets
below $0.01$~mag. Hence, we can safely discard misalignments as a
possible source of systematic uncertainty in the color
gradients. Another source of error may come from the estimates of the
background. The v1.0 present a non-negligible background which was
carefully estimated and subtracted in our analysis (see
\S\S6.2). Furthermore, we checked the possible effect of a bad
background estimate on the color gradients by re-running our analysis
on the sample with either no background correction or twice of the
original estimate in each band. No significant change in the color
gradients was found.

\subsection{Adaptive Binning}

A pixel-by-pixel analysis is an obvious avenue to pursue in the
study of unresolved stellar populations (Menanteau, Abraham \& Ellis 2001).
However, the steepness of the surface brightness profile of 
early-type galaxies implies the pixel distribution spans a wide 
range of signal-to-noise ratios (SNR). Adaptive binning is thereby preferred.
One common binning procedure reduces the color information
to a one dimensional surface brightness profile done by adding the
light inside elliptical annuli. At the cost of smoothing out the 
color distribution, this method gets the maximum SNR from the data.

\begin{figure}
\plotone{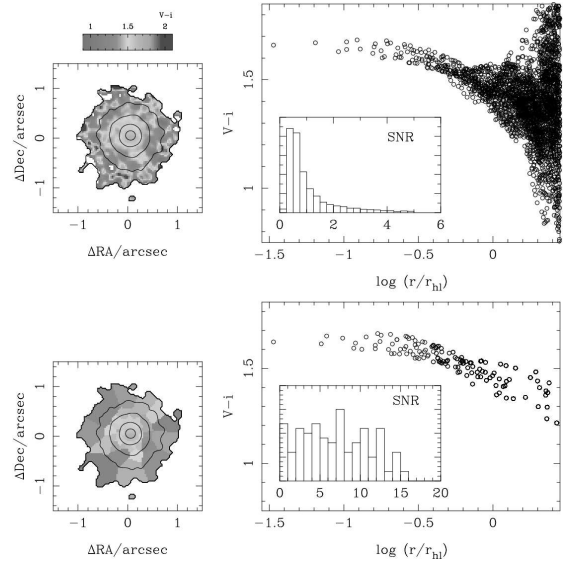}
\caption{Pixel-by-Pixel and Voronoi tessellation
analysis of a sample early-type galaxy: GDS J$033246.05-275444.8$ 
($i_{\rm AB}=22.14$, $z=0.64$). 
The top panels show the results for a pixel by pixel analysis. The image
gives the $V-i$ color map and the contours track the surface brightness
in the $i$-band from 21 to 24 in steps of 1~mag/arcsec$^2$. The bottom panels
show the analysis on an adaptive binning following an Voronoi Tessellation
as described in the text. The target SNR in the flux ratio $\Phi (V)/\Phi (i)$ 
is $10$, corresponding to
a photometric error in $V-i$ of $\sim 0.1$~mag. The right panels show the
color profile and the histogram of SNR in either case. 
\label{fig:voro}}
\end{figure}

Adaptive binning is a better option as it preserves as 
much of the spatial information as possible. Using a tree 
structure is the first choice, which starts with the 
full image of the galaxy as the root and then subdivides it into
smaller squares aligned with the axes with the criterion of 
further subdivisions based on the SNR measured in each square.
However, this method is only effective if the observed 2D 
distribution is smooth enough. 
Furthermore, the scatter of the SNR about the target value is far from
optimal in most cases. Cappellari \& Copin (2003) presented an
optimal binning algorithm based on a Voronoi tessellation. 
A local density is defined as the square of the SNR, which reduces the binning 
algorithm to finding a tessellation that encloses equal masses 
according to this density. The algorithm includes a morphological
criterion in order to generate bins as round as possible, an important
feature in our analysis as filamentary bins would greatly distort
the color gradients.
We decided to base our color analysis on a Voronoi tessellation
with a target SNR of 10 for the $B/i$, $V/i$ or $i/z$ flux ratios. 
This implies a homogeneous photometric uncertainty of 
$\sim 0.1$~mag in the colors.

\begin{figure*}
\centerline{\epsfxsize=\hsize{\epsfbox{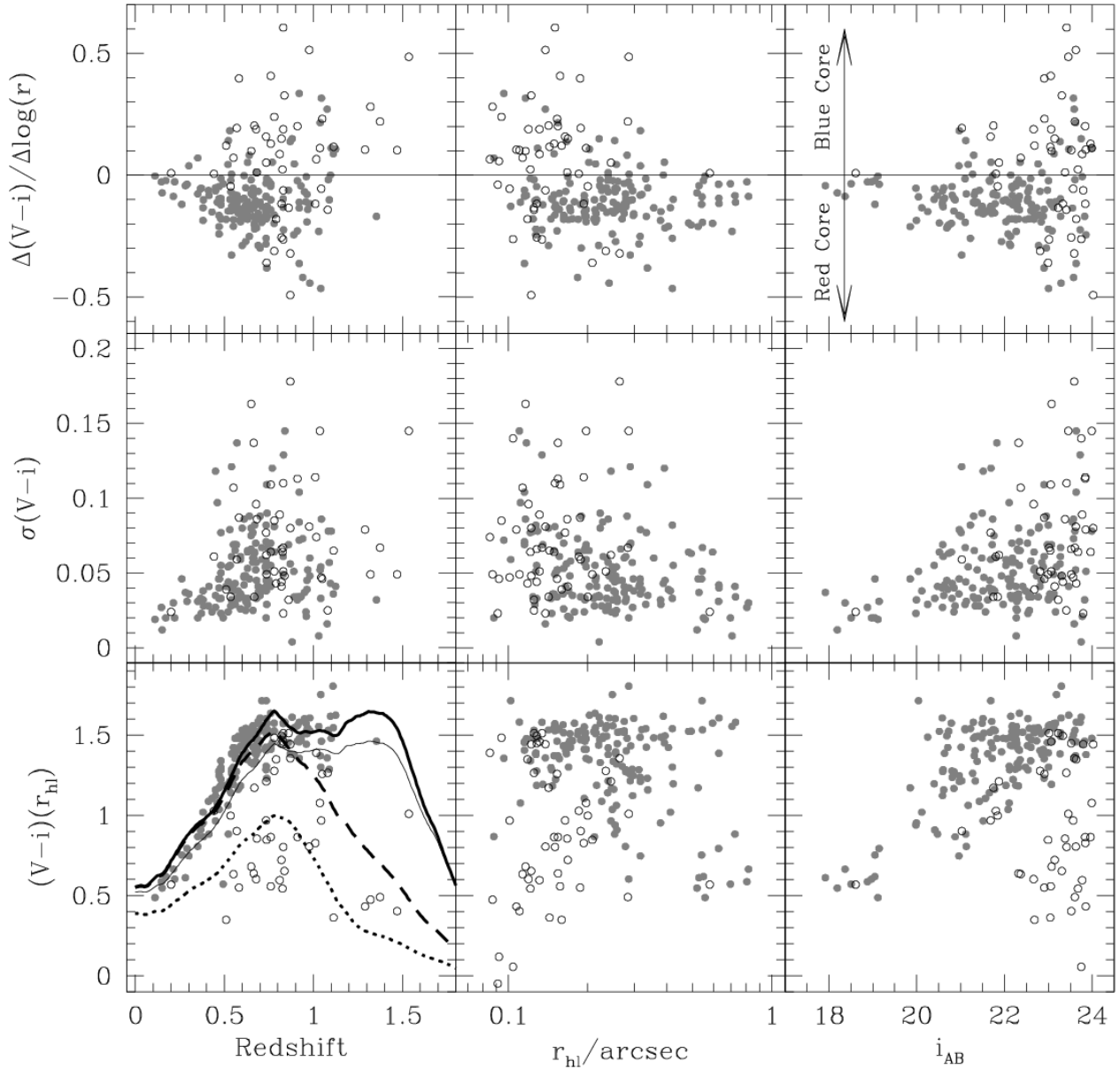}}}
\caption{$V-i$ color distribution. From top to bottom, the radial color 
slope, scatter and color at the half-ligh radius of our sample of 
early-type galaxies is shown
as a function of redshift ({\sl left}); half-light radius ({\sl middle});
and $i$-band apparent magnitude ({\sl right}). The sample is split
into those with a photometric type earlier (filled dots) or later 
(hollow dots) than $1\frac{1}{3}$, which marks the boundary between
our selection of red and blue systems. The thick lines in the
bottom-left panel give the color evolution of a stellar population 
with solar metallicity and an exponentially decaying star formation rate,
started at $z_F=5$ and with a decay timescale of 0.5 (solid), 
1 (dashed) and 8~Gyr (dotted). The thin solid line shows the
evolution with a timescale of 0.5~Gyr at 1/3 of solar metallicity.
\label{fig:Vi}}
\end{figure*}

Figure~\ref{fig:voro} illustrates the advantage of an adaptive binning
technique with respect to a pixel-by-pixel analysis. We compare these
two methods with a $i_{\rm AB}\sim 22$ galaxy at $z\sim 0.6$ 
(GDS J$033246.05-275444.8$). The top panels show the $V-i$ color map
(left) and color-radius distribution (right) along with the distribution
of the SNR (inset). The histogram shows a wide distribution with
a large number of pixels with SNR$\lesssim 1$. Such a distribution 
gives a wide
scatter of colors in the outer parts of the galaxy. The bottom panels
show the same galaxy after an optimal Voronoi tessellation with a 
target SNR$=10$. The color map appears much smoother but not loosing
real inhomogeneous color distributions. The histogram of SNR 
shows that there is a very significant reduction in the number of
(binned) pixels with a low SNR. The analysis of the color-radius 
distribution is thereby more robust and less prone to systematic
effects from the noise.

\subsection{Color gradients and scatter}

For each target, we define the 2D region corresponding to the galaxy 
from the SExtractor segmentation maps, defined at the $1\sigma$
flux level in the deeper $i$ band. The background is estimated
in each bandpass by taking the median of the pixels not belonging
to the enlarged segmentation map of any source inside a $701\times 701$ 
pixel region ($21\times 21$~arcsec$^2$) centered at the target. 
After background subtraction, the resulting image was convolved with 
the point spread function (PSF) of the other passband used in the 
analysis. Degrading both images to a common resolution is more robust
than deconvolving the images by their PSFs.
Deconvolution introduces noise and its 
effect on the final color maps is hard to assess. We explored two 
different sets of PSFs, stellar ones from bright stars free of
nearby sources and synthetic ones using Tiny~Tim (Krist 1993),
see appendix for details.

\begin{figure}
\plotone{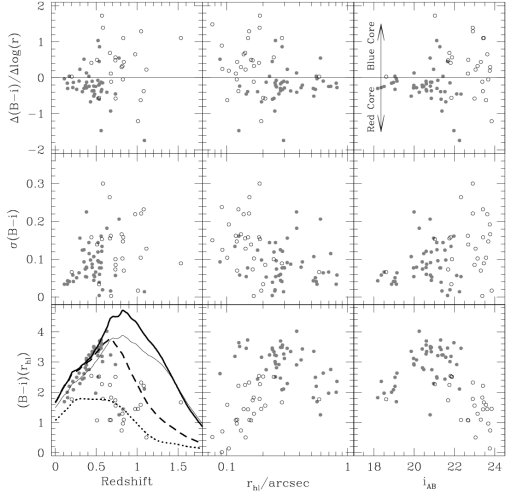}
\caption{Same as figure~\ref{fig:Vi} for $B-i$ color.
\label{fig:Bi}}
\end{figure}

Furthermore, in order to minimize a possible systematic effect 
from the PSF, we decided to reject pixels inside $r=0.1$~arcsec 
from the galaxy centre. These
are the pixels most affected by the smearing effect of the PSF,
and the simulations show that such a truncation
yields results which are insensitive to the choice of PSF.
The results presented in this paper are based on a convolution
with the -- more realistic -- stellar PSFs, although a comparison
with the second analysis based on the synthetic PSFs indeed show
that the color zero points, gradients and scatter do not
change significantly. 

After the convolution, the image is binned 
according to a Voronoi tessellation as described above. A linear
fit is then applied to the resulting bins as a function of 
$\log (r/r_{\rm hl})$, where the half-light radius is measured 
from the Petrosian radius. The fitting method has to be robust
not to give much weight to the outliers. Hence, we start fitting the
color distribution with an M-estimator based on the mean absolute
deviate (Press et al. 1992). We cull the outliers which fall
outside of a band extending 4 times the measured scatter with respect
to this fit, after which we apply a least-squares fit to a reduced set 
without these outliers. This result gives the slope and zero point 
used in our analysis. Finally, the scatter is computed about this
linear fit for the full data set.
For each galaxy, we thereby come up with three numbers to describe 
the color distribution,
namely zero point (color at $r=r_{hl}$), radial color slope 
($\Delta$ Color$/\Delta\log r$) and scatter (about the best linear fit
to the color radial distribution). The scatter is
obtained from the biweight measure of scale (Beers, Flynn \& Gebhardt 1990)
which is much more robust with respect to the presence of outliers.
Nevertheless, we also computed the RMS scatter and found no
significant discrepancies between these two estimates of
the scatter. 

\begin{figure}
\plotone{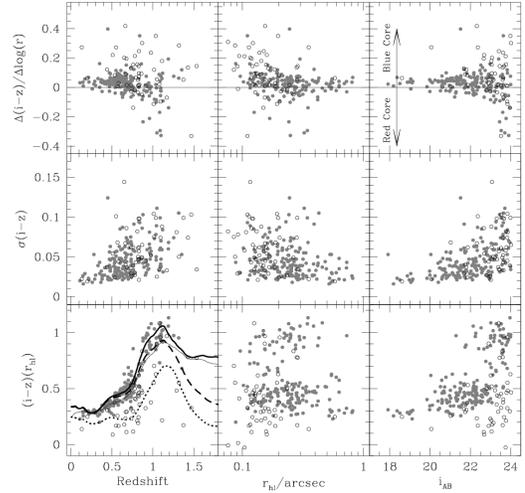}
\caption{Same as figure~\ref{fig:Vi} for $i-z$ color.
\label{fig:iz}}
\end{figure}

The results for the observed color distribution are shown in
figures~\ref{fig:Vi}, \ref{fig:Bi} and \ref{fig:iz} for $V-i$, $B-i$,
and $i-z$, respectively.  The top, middle and bottom panels of the
three figures show the color gradient, scatter and color at $r=r_{hl}$
as a function of redshift ({\sl left}), angular size ({\sl middle}) or
$i_{\rm AB}$ magnitude ({\sl right}). The sample is subdivided into
red (filled dots) and blue (hollow dots) galaxies, as described above.
The bottom panels of figures~\ref{fig:Vi}, \ref{fig:Bi} and
\ref{fig:iz} include the color-redshift evolution of a simple stellar
population (Bruzual \& Charlot 2003 models; Salpeter IMF) with solar
metallicity and formation redshift of $z_F=5$, with an exponentially
decaying star formation rate $\psi (t)\propto\exp (-t/\tau_{SF})$ with
timescales $\tau_{SF}=0.5$ (solid), $1$ (dashed) and $8$~Gyr
(dotted). The thin line corresponds to $\tau_{SF}=0.5$~Gyr and
metallicity $Z_\odot /3$.  Notice the remarkable segregation with
respect to photometric type. The color segregation should not come as
a surprise since the colors themselves are the basis for the
photometric classification. However, this segregation is followed by
the color gradients, so that objects with blue cores -- positive
radial color gradients -- belong to the late photometric
type. Similarly, most of the red objects are found with red
cores. Such a behavior is {\sl suggestive} of an inside-out buildup of
the stellar component, i.e. all recent star formation (in blue
galaxies) takes place at the core, whereas the red galaxies have their
oldest populations close to the center.  In principle, one could have
expected an increased scatter in bluer galaxies -- caused by recent
star formation. However, the figures do not show such segregation with
respect to color, and only display the expected increase in the
scatter towards fainter galaxies. This segregation is seen in all
three HST/ACS colors (see the top panels of figures~\ref{fig:Vi},
\ref{fig:Bi} and \ref{fig:iz}).

\begin{figure}
\plotone{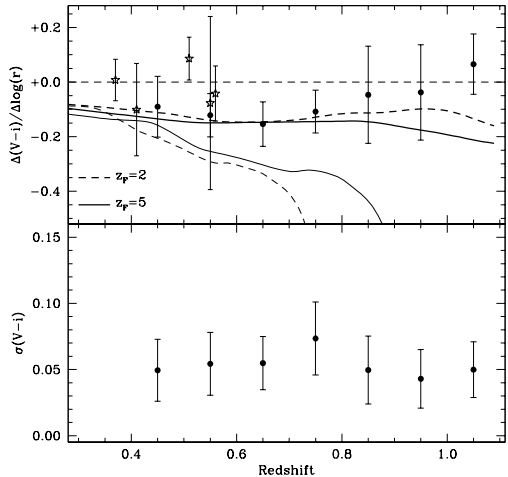}
\caption{$V-i$ gradients ({\sl top}) and scatter ({\sl bottom}) 
as a function of redshift (only red early-types considered in this figure). 
The stars are cluster data
from Tamura \& Ohta (2000). Two alternative evolution models are considered depending
on whether the color gradient is caused by metallicity (thick lines) or by age
(thin lines). Two formation redshifts are shown for each case: $z_F=2$ (dashed lines)
and $z_F=5$ (solid lines).
\label{fig:grads}}
\end{figure}

In figure~\ref{fig:grads} we present the data in redshift bins, showing the
evolution of the color gradient ({\sl top}) and scatter ({\sl bottom}).
The stars are cluster data from Tamura \& Ohta (2000). We have also included
two simple models similar to those presented in Tamura et al. (2000).
These two alternative scenarios invoke either a pure age (thin lines)
or a pure metallicity radial gradient (thick lines). 
We used the population synthesis models of Bruzual \& Charlot (2003)
with a Salpeter IMF.
The models are constrained at $z=0$ by the mean value of 
$\Delta (B-R)/\Delta\log r=-0.09$~mag/dex from Peletier et al. (1990).
For each model we choose two possible ages for the stellar populations at 
the center: $z_F=2$ (dashed) and $z_F=5$ (solid lines). 

The data are compatible
with a pure metallicity radial gradient and rules out a significant spread
in stellar ages within the galaxy. Furthermore, no significant 
difference is found between our field sample and the cluster galaxies
from Tamura \& Ohta (2000). The rise in the color gradient at $z\sim 1$
is caused by the larger number of galaxies with blue cores.
The lower panel shows the redshift evolution of the scatter, which can
be used as a proxy for patchy star formation. No significant change 
is found with redshift, in support of a quiescent evolution of the stellar 
populations. The trend found in figure~\ref{fig:grads} imposes powerful 
constraints
on the formation process of elliptical galaxies, suggesting an
early and short-lived formation epoch followed by passive evolution. 
The galaxies which are undergoing some residual star formation
at those epochs are mostly faint and small, supporting a downsizing
picture of galaxy formation. The lack of features in the color gradients
and the small scatter furthermore suggest that $z<2$ mergers which lead
to early-type galaxies must be essentially dissipationless events.

\section{Discussion}

An important aspect of our work is that, by construction, we investigate
the intermediate redshift population of early-type galaxies which, at those
epochs, already look like today's elliptical galaxies. This approach has 
been followed in other studies of the massive galaxy population at moderate
redshift, e.g. to study the color-magnitude relation (Stanford, Eisenhardt
\& Dickinson 1998) and the fundamental plane (Wuyts et al. 2004; van~der~Wel
et al. 2004). Van~Dokkum \& Franx (2001) have discussed this issue, 
christening it the ``progenitor bias'' as any star-forming galaxy
at those epochs -- which may nonetheless become part of an 
elliptical at $z=0$ -- would have been excluded from such analyses.
In this respect, we note that the search for intermediate-$z$ red
galaxies {\it without} an early-type morphology that we have conducted
on the GOODS/CDFS dataset returns only 30 objects at $i_{\rm AB}\leq 24$.
Spitzer and ALMA will help assessing the fraction of dust-enshrouded
galaxies which are compatible with their evolution into $z=0$
early-type galaxies. While ellipticals which are built by a relatively late
merger are not accounted in our own and the other quoted investigations, 
studying the properties of an intermediate-$z$ morphologically-selected
sample of ellipticals allows us to explore important issues pertaining
to the evolution of massive galaxies. Specifically, the morphological 
approach allows us not only to set a lower limit on the fraction of
the $z=0$ early-type population which is already in place at $z\sim 1$,
but also to investigate whether star formation proceeds in these systems
well after their mass assembly has occurred. This is an important 
constraint to current galaxy formation models.

\begin{figure}
\plotone{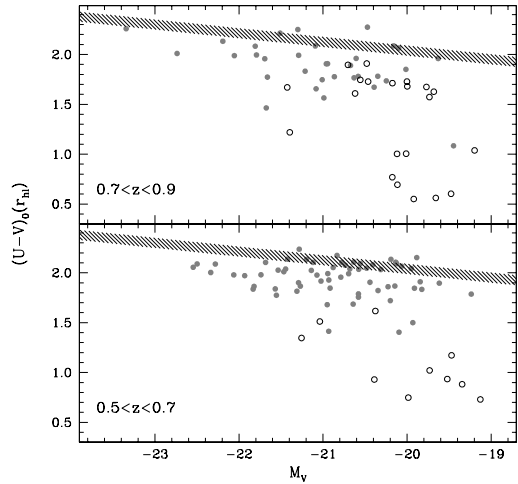}
\caption{The color-magnitude relation in rest frame $U-V$.  The shaded
area represents the local color-magnitude relation of the Virgo or
Coma clusters from Bower et al. (1992). Higher- (top) and
lower-redshift subsamples are shown, split at the median redshift,
with a redshift interval ($\Delta z=0.2$) chosen not too large to
avoid the scatter from the uncertainties in the K-correction. Filled
and hollow circles correspond to our photometric selection of red and
blue galaxies, respectively (see text for details).
\label{fig:uv0}}
\end{figure}

Figure~\ref{fig:uv0} shows the rest frame $U-V$ color magnitude relation
for two subsamples separated in redshift with respect to the median.
A small redshift interval $\Delta z=0.2$ has been chosen in order to 
reduce the scatter from the uncertainties in the K correction.
The shaded area is the local CMR of the Coma or Virgo cluster 
(Bower et al. 1992) in the AB system.
Notice that practically all red galaxies
follow the red sequence except for a systematic blueing caused by the
lookback time, as seen in the early-types of clusters at moderate
and high redshift (Stanford et al. 1998; Blakeslee et al. 2003). 
Most of the blue galaxies fall outside of this red sequence,
and preferentially represent lower mass systems at higher redshift.

We compare our sample with a simple
model which assumes a two-burst formation scenario. Each burst is
characterized by an exponentially decaying star formation rate 
and is formed at times $t(z_F)$ (old) and $t_Y$ (young), respectively:
\begin{equation}
\psi(t) \propto f_Y e^{-(t-t_Y)/\tau_{SF}} + 
(1-f_Y) e^{-[t-t(z_F)]/\tau_{SF}},
\end{equation}
where the star formation timescale is $\tau_{SF} =0.5$~Gyr, $f_Y$ is the
stellar mass fraction in young stars. The old component is formed 
at $z_F=5$, and $(t-t_Y)\leq 1$~Gyr, i.e. the young component is formed in the
last 1~Gyr. The metallicity is the same for both components but we 
choose a random value uniformly in the range 
$-0.3\leq\log (Z/Z_\odot )\leq+0.3$.

\begin{figure}
\plotone{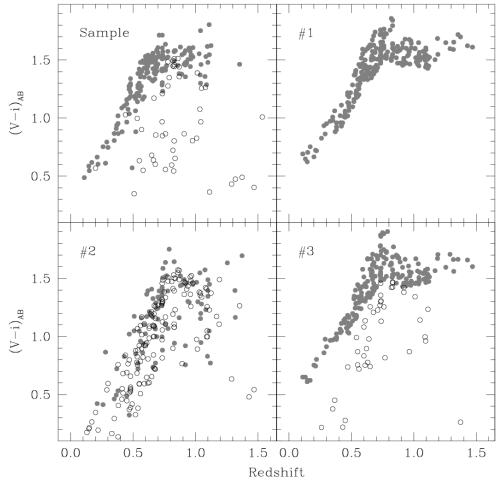}
\caption{Comparison of the observed $V-i$ color vs redshift diagram with 
a simple model with 2~bursts of star formation. The panel on the top left
shows the sample (filled and hollow circles represent red and blue galaxies,
respectively). Model \#1 assumes a single burst at $z_F=5$ with a star formation
timescale $\tau_{SF}=0.5$~Gyr, model \#2 includes a later second burst 
in the last 1~Gyr \emph{for all galaxies}. Model \#3  
imposes this second burst only on 20\% of the galaxies (see text for details).
The synthetic subsamples are coded depending on whether they have a significant 
fraction of young stars (hollow circles; $f_Y\geq 0.2$) or are dominated by 
old stars (filled circles).\label{fig:2burst}}
\end{figure}

Figure~\ref{fig:2burst} compares the $V-i$ color versus redshift relation
with three such models. The top left panel shows our sample. 
The simulated galaxies have the same
redshift distribution as the original sample and are subdivided with respect
to their young stellar mass fractions so that the late-types (hollow dots)
have $f_Y\geq 0.2$. Model \#1 corresponds to 
a single burst ($f_Y=0$) for all galaxies. 
Model \#2 assumes a second burst for all galaxies, randomly chosen in 
the range ($f_Y\leq 0.5$). Finally, model \#3
applies this second burst \emph{only} to 20\% of the galaxies chosen
in a random way. This fraction corresponds to the contribution of 
blue galaxies to our sample. 

As expected, the red galaxies are compatible with a single, early
burst scenario -- which represents most of the galaxies with a red
core. The clear cut separation between the two photometric types can
not be explained by a homogeneous distribution of galaxies with a
second burst. Model \#2 gives a smooth transition between the older
and the younger galaxies. However, model \#3 shows that the data could
be reproduced by a model in which a certain fraction of the galaxies
is undergoing a second star forming episode. Some of these galaxies
could present the spectral features of a post-starburst E+A
galaxy. The fast color evolution along with the finite sampling
reproduces the observed bimodality with respect to color.  The
presence of blue cores in blue galaxies suggest these recent episodes
of star formation take place at the centre.  Furthermore,
figure~\ref{fig:uv0} implies that most of the recent star formation
appears in the fainter galaxies, i.e. ``down-sizing'' (see e.g. Kodama
et al. 2004; de~Lucia et al. 2004; McIntosh et al. 2005).  Such a
trend would be expected from a strong correlation between galaxy mass
and star formation efficiency as proposed by Ferreras \& Silk (2003;
2000a) and is in agreement with various observed properties such as
the large scatter at the faint end of the near-ultraviolet $-$ optical
color-magnitude relation (Ferreras \& Silk 2000b) or the Fundamental
Plane of field early-type galaxies at moderate redshift (van der Wel
et al. 2004).  The ``red sequence'' of model \#3 is compatible with
the data out to the highest redshifts and implies that $\sim$ 95\% of
the stellar mass component in these galaxies should have been formed
$\sim 1.5$~Gyr after $z_F=5$, i.e. at redshifts $z\gtrsim 2.5$.

\section{Summary and Conclusions}

The internal color distribution of a large sample of early-type galaxies 
can be used to constrain the mechanisms leading to the formation
and assembly of their stellar populations. In this paper we explore
a large sample of field early-types with $i_{\rm AB}\leq 24$ from the HST/ACS
images of the GOODS/CDFS field, which cover a solid angle $\sim 160$~arcmin$^2$. 
The classification of early-types starts with  a pre-selection
of the targets using recent quantitative morphology techniques
followed by a visual inspection. A total of $380$ objects 
were selected as early-type galaxies, i.e. with the visual appearance
of an elliptical or a lenticular galaxy. 


\begin{deluxetable*}{cccccccc}
\tabletypesize{\scriptsize}
\tablecaption{The sample of GOODS/CDFS early-type galaxies.
\label{tab:eGoods}}
\tablewidth{0pt}
\tablehead{
\colhead{ID} & \colhead{$i_{\rm AB}$} &
\colhead{$r_{\rm hl}/$arcsec} & t$_{\rm phot}$ & \colhead{$z^a$} &
\colhead{$\langle V-i\rangle_{\rm rhl}$} 
& \colhead{$\Delta (V-i)/\Delta\log r$} & \colhead{Scatter}}
\startdata
$i_{\rm AB} < 23.0$  & & & & & & & \\
\\
 J033157.84-274245.2   & $20.63$ & $0.63$ & $1.00$ & $(0.610)$ & $1.386\pm0.031$ & $-0.095\pm0.052$ & $0.04$  \\
 J033158.13-274459.4   & $21.02$ & $0.29$ & $1.00$ & $(0.540)$ & $1.215\pm0.009$ & $-0.328\pm0.033$ & $0.12$  \\
 J033159.76-274411.4   & $21.51$ & $0.25$ & $1.30$ & $(0.450)$ & $0.965\pm0.031$ & $-0.736\pm0.114$ & $0.12$  \\
 J033202.51-274536.0   & $22.66$ & $0.12$ & $3.00$ & $0.678$ & $0.602\pm0.007$ & $0.188\pm0.063$ & $0.10$  \\
 J033202.71-274310.8   & $18.51$ & $0.70$ & $1.30$ & $0.493$ & $0.571\pm0.001$ & $-0.035\pm0.001$ & $0.02$  \\
 J033206.88-274207.6   & $20.96$ & $0.32$ & $1.30$ & $(0.300)$ & $0.747\pm0.004$ & $0.038\pm0.012$ & $0.04$  \\
 J033208.10-274732.7   & $22.81$ & $0.23$ & $1.70$ & $(0.780)$ & $1.276\pm0.057$ & $-0.311\pm0.233$ & $0.05$  \\
 J033208.41-274231.3   & $22.03$ & $0.13$ & $1.00$ & $(0.480)$ & $1.186\pm0.005$ & $-0.101\pm0.044$ & $0.05$  \\
 J033208.45-274145.9   & $22.18$ & $0.24$ & $1.00$ & $(0.660)$ & $1.558\pm0.030$ & $-0.053\pm0.119$ & $0.06$  \\
 J033208.53-274217.7   & $21.71$ & $0.28$ & $1.00$ & $(0.690)$ & $1.584\pm0.015$ & $-0.157\pm0.058$ & $0.04$  \\
\enddata
\tablecomments{A full version of table \ref{tab:eGoods} is published in its 
entirety in the electronic edition of the {\it Astrophysical Journal}. 
A portion is shown here for guidance regarding its form and content.}

\tablenotetext{a}{Photometric redshifts are enclosed in brackets.}
\end{deluxetable*}

The Kormendy relation is then used
to assess which of the candidates will evolve into present day
ellipticals. The results shown in figure~\ref{fig:kor} clearly suggest
a dual nature for the blue galaxies. Those at the
lower redshifts fall below the local relation, so that -- notwithstanding 
any major structural change -- they will evolve away from the Kormendy 
relation defined by local early-type galaxies. 
These galaxies are spirals with strong
ongoing star formation, or even starbursts. On the other hand, the
high-redshift sample of blue galaxies falls within the
Kormendy relation. We define a final sample 
of {\em bona fide} ellipticals using this criterion.
The sample comprises about one third of
the original sample where most of the galaxies with blue colors 
are at the higher redshifts. The analysis of the number counts 
versus apparent magnitude suggests a significant evolution
of the comoving number density of galaxies with an early-type morphology, 
scaling as $n\propto (1+z)^{-2.5}$.
Similar results have been found using different approaches and
datasets by other independent studies (e.g. Lin et al. 1999; 
Bell et al. 2004). Such an evolution should therefore be considered
as a robust and important constraint to be reproduced by galaxy 
formation models.

For each galaxy in our final sample, we obtain an estimate of 
the zero point, slope 
and scatter of a range of colors from the available photometry
of the ACS/GOODS database ($B,V,i,z$). These colors -- given 
the redshift range of the sample ($z\lesssim 1.5$) -- map a 
rest frame wavelength range which straddles the age-sensitive 
4000\AA\  break. The zero point of each measurement is equivalent 
to the integrated color of the galaxy, and so it traces the bulk 
properties of the stellar
populations. The slope of the radial gradient of the color 
accounts for age or metallicity variations and its evolution with 
redshift is a powerful discriminator among star formation 
histories (Tamura \& Ohta 2000, Kodama \& Arimoto 1997). 
The scatter can be used to track possible small episodes of star 
formation, as suggested by the line strength analysis 
of Trager et al. (2000)

The results are presented in figures~\ref{fig:Vi}, \ref{fig:Bi}
and \ref{fig:iz}, with the remarkable finding of a clear cut segregation
between the colors \emph{and gradients} of the sample. This 
bimodality is strongly correlated with the type
assigned from the photometric redshift estimate of Mobasher et al. (2004).
Such correlation is robust with respect to the effects of the PSF,
the galaxy luminosity or its size. The correlation of photometric
type with color should not come as a surprise. However, the non-trivial
finding is the correlation with color gradient: Galaxies with blue cores
tend to be of later photometric types. A comparison with simple models
of star formation in two bursts suggest that the blue subsample could
represent a population of early-type galaxies which have experienced
a recent episode of star formation in the recent $\sim 1$~Gyr. The
fraction of these galaxies decreases as $z\rightarrow 0$. Furthermore,
figure~\ref{fig:uv0} shows that these galaxies are the fainter ones,
a result which is consistent with a lower star formation efficiency
in low mass early-type galaxies. 
Figure~\ref{fig:grads} presents a strong constraint which should
be imposed on models of galaxy formation. The color gradient
is not found to undergo a strong redshift evolution, and is 
compatible with local early-type galaxies, assuming a radial gradient
of metallicity, but not age. This trend is seen out to $z\sim 1$.
Furthermore, the color scatter does not evolve with redshift.
This result along with the correlation between color and color
gradient suggests an ordered inside-out formation process and the
lack of mergers with star formation at $z\lesssim 2$. 


\acknowledgments 
We would very much like to acknowledge the useful comments and
suggestions from the referee, Dr. Eric Bell.  T.L. acknowledges
support by the Swiss National Science Foundation.  This paper is based
on observations made with the NASA ESA \emph{Hubble Space Telescope},
obtained from the data archive at the Space Telescope Science
Institute, which is operated by the Association of Universities for
Research in Astronomy, Inc.  under NASA contract NAS 5-26555. The HST
ACS observations are associated with proposals 9425 and 9583 (the
GOODS public imaging survey).



\begin{figure}
\plottwo{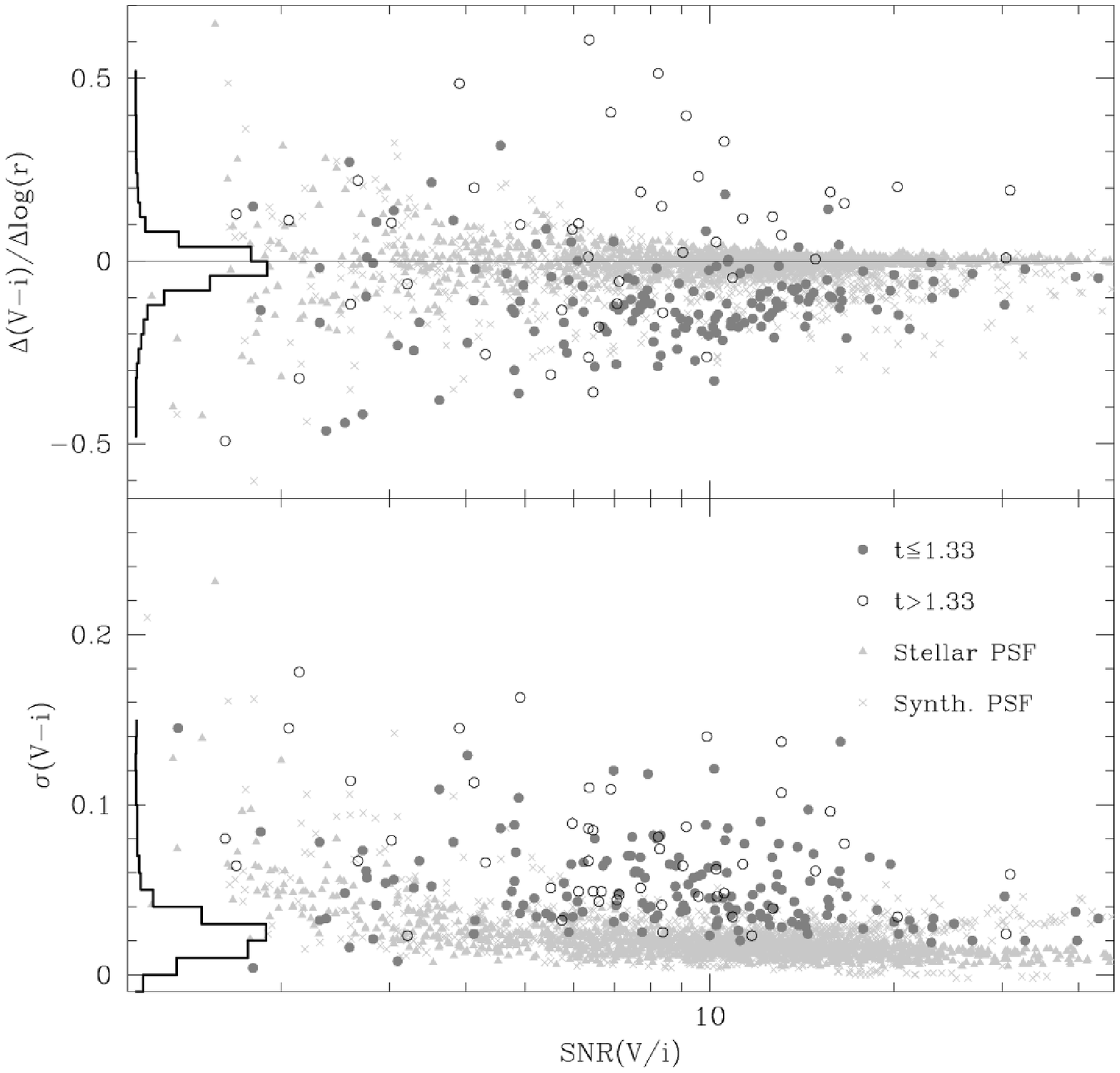}{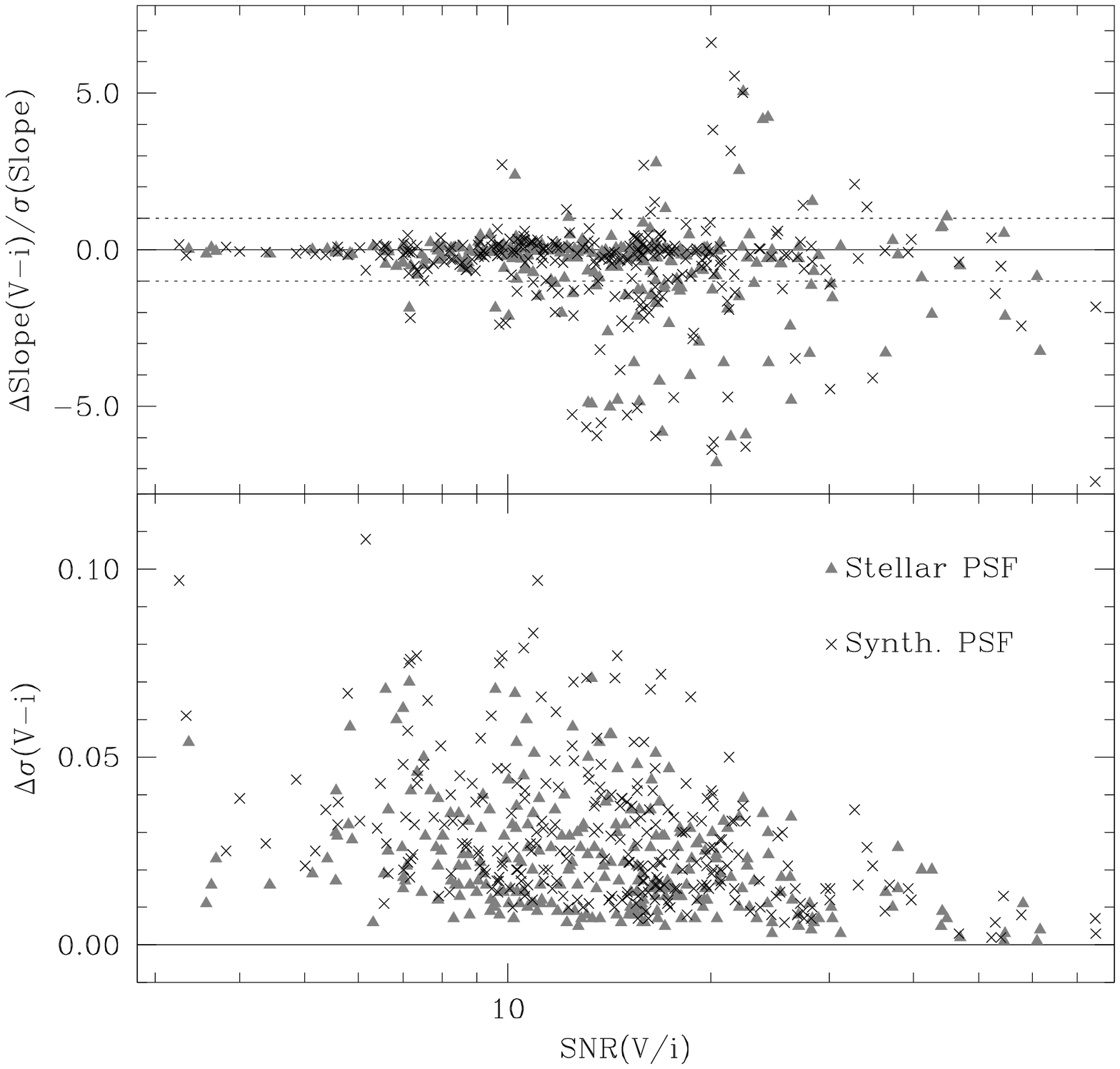}
\caption{{\sl Left:} Comparison of the $V-i$ radial slope and scatter
as shown in figure~\ref{fig:Vi} with respect to a synthetic sample of
$>1,500$ galaxies with zero color gradient and with the same noise
behavior as the ACS/GOODS data: triangles (stellar PSF used in the
analysis) and dots (Tiny-Tim PSF). {\sl Right: } Recovery of the
observed slopes in a sample of synthetic images with the same apparent
magnitudes, sizes and noise characteristics as the original
sample. Two different PSFs were used for the analysis, a stellar PSF
as the one used in our analysis (triangles) and a synthetic one from
Tiny-Tim (crosses).
\label{fig:simul}}
\end{figure}

\appendix

\section{The effect of the SNR on color gradients and scatter}

We performed a set of simulations in order to assess the uncertainties
related to the observed range of signal to noise ratios as well as the
effect of the PSF and the pixel sampling. Four sets of galaxies each
were generated, with a smooth surface brightness corresponding to a
S\'ersic profile with an index between $3-4$.  Two tests were
performed. In the first one we imposed a zero color gradient and no
intrinsic scatter. We chose a random distribution of ellipticities and
orientations, but the magnitudes and sizes corresponded to those from
our sample. Each synthetic set thereby comprises $370$ galaxies with
an identical distribution of sizes and magnitudes as those of the real
sample. In order to increase the statistical significance of the
analysis we ran 4 such samples, for a total of $1,480$ galaxies. Noise
was subsequently added to the image with the same distribution of SNR
as a function of surface brightness as those obtained from the images
and weight maps of the HST/ACS GOODS frames. Finally, we convolved the
synthetic images with two choices of PSF, namely stellar -- obtained
from stars in the GOODS images with a high SNR --- and synthetic --
from TinyTim (Krist 1993).  The resulting images followed the same
pipeline analysis as the real galaxy sample. The results are shown in
figure~\ref{fig:simul} ({\sl left}), where the $V-i$ slope ({\sl top})
and scatter ({\sl bottom}) is shown as a function of SNR. The real
data is overplotted -- solid and hollow circles representing our
sample of red and blue galaxies, respectively.  The synthetic data are
shown as triangles (stellar PSF) or crosses (synthetic PSF). One can
see the increasing noise in the estimated slope and scatter towards
lower SNR.  The figure shows that the red/blue core separation as a
function of photometric type is intrinsic to the galaxies.

We performed a second test on a synthetic sample with the same size,
apparent magnitude {\sl and color gradient} as our real sample. The
procedure is equivalent as the one described above, and we show the
results in figure~\ref{fig:simul} ({\sl right}), where we plot in the
top panel the difference between imposed and measured slope divided by
the uncertainty. Notice that the quantity $\Delta {\rm
Slope}(V-i)/\sigma ({\rm Slope})$ has a mean around zero and unit
variance ($-0.27$ and $0.90$, respectively, in the simulations). The
bottom panel shows the measured scatter (the synthetic images use a
smooth surface brightness profile but are affected by the noise).
Notice the apparent larger scatter towards high SNR in the top panel.
This is caused by an underestimate of the error bars in brighter
galaxies. The actual difference between imposed and retrieved slopes
{\sl decreases} with increasing SNR. The number of outliers is
consistent with the expected fraction.  Inside the $1\sigma$
confidence level (dotted lines) we have $\sim 70$\% of the complete
sample.

\end{document}